# Enhanced Trustworthy and High-Quality Information Retrieval System for Web Search Engines

Sumalatha Ramachandran, Sujaya Paulraj, Sharon Joseph and Vetriselvi Ramaraj

Department of Information Technology, MIT, Anna University,
Chennai - 600044, India.

**Abstract**
The WWW is the most important source of information. But, there is no guarantee for information correctness and lots of conflicting information is retrieved by the search engines and the quality of provided information also varies from low quality to high quality. We provide enhanced trustworthiness in both specific (entity) and broad (content) queries in web searching. The filtering of trustworthiness is based on 5 factors – Provenance, Authority, Age, Popularity, and Related Links. The trustworthiness is calculated based on these 5 factors and it is stored thereby increasing the performance in retrieving trustworthy websites. The calculated trustworthiness is stored only for static websites. Quality is provided based on policies selected by the user. Quality based ranking of retrieved trusted information is provided using WIQA (Web Information Quality Assessment) Framework.

## 1. Introduction

Information comes from increasingly diverse sources of varying quality. There is no guarantee for the correctness of information on the Web. Also, different websites often provide conflicting information [8], as shown in the following example.

Example: (Runtime of Harry Potter 6 movie)
The user gives an input query as "Harry Potter 6 runtime" in the Google Search Engine. The information retrieved is as follows: 8 websites gave the information as 153 mins, 7 websites as 138 mins, 2 websites as 144 mins, 1 website each as 94 mins and 104 mins. Thus it is clear that the search results are not correlative and credible [9].

Information quality is task-dependent. A user might consider the quality of a piece of information appropriate for one task but not sufficient for another task. Information quality is subjective, as a second less quality concerned user might consider the quality of the same piece of information appropriate for both tasks. Which quality dimensions are relevant and which levels of quality are required for each dimension is determined by the specific task at hand and the subjective preferences of the information consumer.

## 2. Related Works

A framework for the Veracity problem is given in [1] i.e., Conformity to truth, which studies how to find true facts from a large amount of conflicting information on many subjects that is provided by various websites. This framework helps us to find trustable websites and true facts. An algorithm called TRUTHFINDER is existing in [1] for the Veracity problem, which utilizes the relationships between websites and their information. A website is trustworthy if it provides many pieces of true information and a piece of information is likely to be true if it is provided by many trustworthy websites. An iterative method is used to infer the trustworthiness of websites and the correctness of information from each other. For selecting trustworthy information, the TRUTHFINDER uses two parameters – Website trustworthiness and Fact confidence. The limitations in TRUTHFINDER are that, the initial assumption of Website Trustworthiness is taken as 0.9 in all cases like popular, authoritative and untrustworthy websites. Only for specific queries (entities) trustworthy websites are retrieved based on single object or property (EX: height of Mt.Everest). Also, Recalculation of trustworthiness of websites for each query given by the user reduces the performance of the system.

The quality of the search results from the web search engines varies as information providers have different levels of knowledge and different intentions [2]. Users of web-based systems are therefore confronted with the increasingly difficult task of selecting high quality information from the vast amount of web-accessible information. In this existing work, the authors introduce the WIQA—Information Quality Assessment Framework**.** The framework enables information consumers to apply a wide range of policies to filter information. The framework employs the Named Graphs data model for the representation of information together with quality related meta-information. The framework uses the WIQA-PL policy language for expressing information filtering policies against this data model. WIQA-PL policies are expressed in the form of graph patterns and filter conditions. The WIQA framework is incorporated into an application called WIQA browser. Implementing this framework in search engines personalizes the web search and helps in the retrieval of high quality information [3]. This is not done in the existing work.

AQUAINT relies heavily on the detection and analysis of design aspects in order to distinguish high quality from low quality pages [6]. But this paper uses subjective policy selection method to assess the quality of the web page.

## 3. Proposed work

In the retrieval of Trustworthy and High-Quality information from web search engines, this paper deals with having Content-Trust in the retrieved web search results.

Content-trust in broad queries based on factors like [5]
- Authority – domain specific,
- Related resources – links from trusted websites
- Popularity- most visited websites,
- Provenance - origin of information provider
- Age - lifespan of time-dependent information

To clear out the assumption made by the TRUTHFINDER, the provenance information about the various websites is used and the initial website trustworthiness is calculated accordingly. The WIQA framework is implemented in web search engines for retrieval of High-Quality and Trustworthy information. Recalculation of website trustworthiness for each query given by the user reduces the performance of the search engine. Hence a method to store the recently calculated trustworthiness in files is provided to improve the search engine's performance [4]. This method is followed only for static websites. Since the content of web 2.0 sites are subject to frequent changes by the users, trustworthiness is not stored in the case of web 2.0 sites.

## 4. Architecture

The proposed overall architecture for the Trustworthy and High-Quality Information Retrieval System for web search engines is given in the following figure – Fig.1.

The user gives the search query in the web search engine interface and also chooses from the various quality policies provided. The search query is given to the searcher which fetches the result URLs from the database.

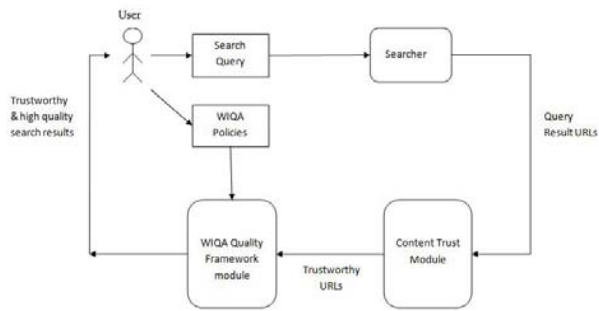

Fig. 1 Trustworthy and High-Quality Information Retrieval System for Web Search Engines

The resultant URLs from the search engine are given as input to the Content Trust Module. The architecture of the Content Trust Module is shown in Fig.2. This Module calculates the website trustworthiness and fact confidence for each of the URLs using the Algorithm for Content Trust Module. The URLs are then filtered based on the calculated trustworthiness. The filtered URLs are given to the WIQA Quality Module which sorts the URLs based on 3 factors – Content, Context and Rating [2]. The Trustworthy and High-Quality Information thus obtained is displayed to the user.

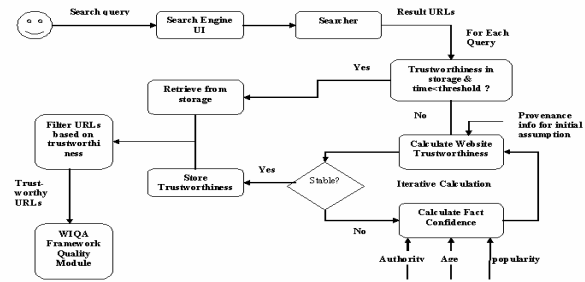

Fig. 2 Content Trust Module

### 4.1 Architecture of Content Trust Module

The search query given by the user is given to the search engine UI. The searcher gives the resultant URLs. For each URL in the result and the trustworthiness given by ω(t) is stored and the time of storage is less than a threshold value, then the value of ω(t) is retrieved from storage. Else, ω(t) is calculated in an iterative way by assuming initial trustworthiness of the website from the provenance information. From this initial value the fact confidence is calculated based on the parameters Authority, Related Resources, Popularity and Age. From this fact confidence, the website trustworthiness is calculated again.

This process continues in an iterative way till ω(t) becomes stable. (i.e., The change in ω(t) compared to the previously calculated ω(t) is minimum). When ω(t) becomes stable, it is stored. URLs are filtered based on the value of ω(t). The filtered URLs are given to the Quality Module.

### 4.1.1 Calculation of Content Trust parameters

The Authority parameter is calculated by analyzing the URL. Different weights are assigned depending on the domain names. The Age parameter is calculated using the metadata like Last-Modified date. A comparison is done with the current date before assigning the value. The Popularity parameter is calculated using the number of In links, which refer to the number of times a particular website is referenced from other trustworthy websites. The Related

Links parameter is used by adding an appropriate weight to each URL's trustworthiness that is listed out in a Highly-Trustworthy website. These 4 parameters are used to calculate the website trustworthiness.

### 4.2 Architecture of Quality Module

The Fig.3. Shows how the Quality is achieved using policies selected by the user. Three quality indicators are chosen in the WIQA Policy Framework. The three quality indicators chosen are information content, contextual information and ratings of the website. Various quality dimensions like Accuracy, Timeliness, Relevancy, Objectivity, Believability etc [2] are measured for each URL. The dimensions which are associated with each quality indicator, depends upon the policy selected by the user.

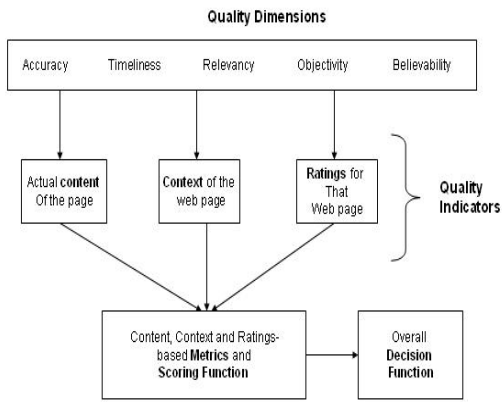

Fig. 3 Achieving Quality

The overall architecture of the Quality Module is shown in Fig.4. The output of the Content Trust module is the input for the Quality module. Trustworthy URLs from the Content Trust module is ranked based on the user selected quality policies.

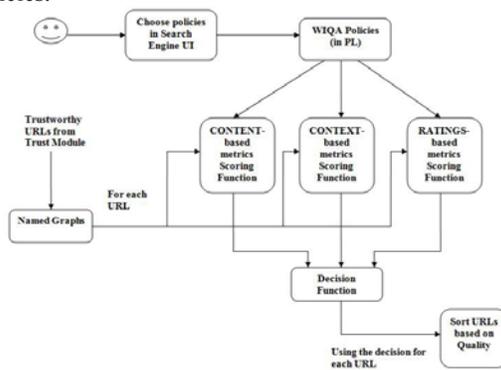

WIQA – web information quality assessment
PL – policy language

Fig. 4 WIQA Framework Quality Module

The user chooses the WIQA policies from the search engine UI. The WIQA policies are written in PL language. The user-selected policies specify the type of task at hand of the user. This is used in the calculation of scoring and decision function.

Information quality assessment metrics can be classified into three categories according to the type of information that is used as quality indicator: (1) information content itself; (2) information about the context in which information was claimed; (3) ratings about information itself or the information provider, as seen in Fig.3.

Based on the three quality indicators chosen by the WIQA Framework, three quality assessment metrics are classified: 1) Content based metric 2) Context based metric 3) Ratings based metric.

Context based metrics are assessed based on the provenance information taken from the metadata. Content based metrics are assessed using the information content itself. Ratings based metrics are assessed by using the rating information taken from the linked RDF data sets for all the websites.

Each of these assessment metrics are used to assess the quality dimensions that are relevant for the task at hand. Based on this, each assessment metric specifies a scoring function to calculate an assessment score. Then a decision function weights assessment scores depending on the relevance of the different assessment metric for the task at hand. The URLs are retrieved from the named graph [7] to be used by the scoring function to calculate assessment scores. Then the URLs are ranked based on the decision function.

## 5. Algorithm for Content Trust Module

The following Content Trust Algorithm calculates the Website Trustworthiness ($\omega(t)$) and Fact Confidence($f(\delta)$) in an iterative fashion where $\delta$ represents an URL.

```
 I/P: Set of URLs
O/P: Trustworthy set of URLs

for(each url[i]) /* i – an integer from 1 to n (no of URLs) */
    {

    /*Initial Website Trustworthiness ω(t)*/

    if(url[i] in <url,trust,time,flag> && flag==0 &&
time<threshold)
      {
       ω(ti)=trust in <url,trust>;
       goto sort;
      }

    else if(url[i] in <url,trust,time,flag> && flag==1)
```

```
    {
     prov=calculate initial ω(tᵢ); /*Retrieving provenance info*/
     ω(tᵢ)=trust+prov;  /*calculating initial assumption for ω(tᵢ)*/
    }
    else
    {
     prov=calculate initial ω(tᵢ); /*Retrieving provenance info*/
     ω(tᵢ)=prov;   /*calculating initial assumption for wt*/
  }

    /*Calculate Fact Confidence*/
    f₁(δ)=retrieveauth(url[i]); /*Retrieve authority info from domain names*/
    f₂(δ)=retrieveage(url[i]); /*Retrieve age info from last-modified date*/
    f₃(δ)=retrievepop(url[i]); /*Retrieve popularity from websites*/
    rel=retrieverel(url[i]); /*Retrieve related websites from outlinks*/

    do               /*Iterative computation*/
    {
    temp= ω(tᵢ);   /* calculation of fact confidence*/

    s(δ) = ∑ₓ₌₁,₂,₃ fₓ(δ);      /*Fact score*/

    if(s(δ)<1)     /* fact confidence */
        f(δ) = log [ -ln (s(δ)) + ω(tᵢ)];
    else
        f(δ) = log [ln (s(δ)) + ω(tᵢ)];

    ω(tᵢ) = e^f(δ) + e^f(δ)/2 ;    /* calculation of website trustworthiness */

    diff= |temp- ω(tᵢ)|;

    } while (diff>0.05);

 trust[i] = ω(tᵢ);

 create entry and store (url,ω(tᵢ),time,0) in <url,trust,time,flag>;

 for(each url in rel)
 {
    if(url in <url,trust,time,flag>)
        break;
    else
    {
        create entry <url,trust,time,flag>;
        trust= ω(tᵢ)/5;
        flag=1;
    }
  }
 }
}
Sort:
sort URLs based on trust[i];
}
```

### 5.1 Formulation:

**Fact Score (s(δ)) :**

$$s(\delta) = \sum_{x=1,2,3} f_x(\delta)$$

**Fact confidence (f(δ)) :**

$$f(\delta) = \log [\ln (s(\delta)) + \omega(t_i)]$$

**Website Trustworthiness (ω(tᵢ)) :**

$$\omega(t_i) = e^{f(\delta)} + e^{f(\delta)/2}$$

$f_x(\delta)$– Weight of the parameters Age, Authority and Popularity

## 6. Result analysis

The preliminary results of the proposed Trustworthy and High-Quality Information Retrieval System for Web Search Engines is shown in Fig.5 and Fig.6

This work uses the open source search engine NUTCH and it is modified to incorporate the Content Trust Module and Quality Module.

Fig.5 shows the unmodified NUTCH search results for the query "Study in US". Fig.6 shows the modified NUTCH search results for the same query.

The search results differ in a way that the modified Nutch provides more authoritative websites like .gov and .edu, more recently modified websites, websites from a trusted source and websites with high popularity in the top search results.

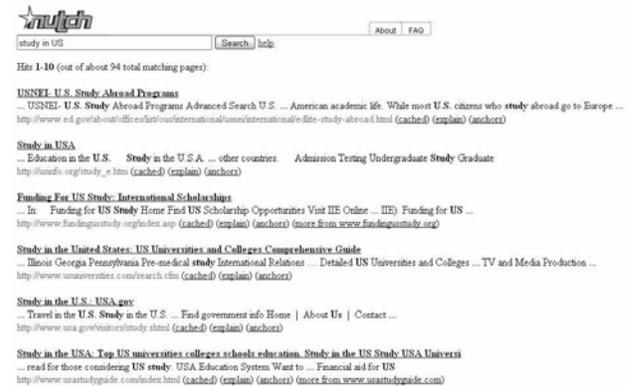

Fig.5. Unmodified NUTCH search results for the query "Study in US"

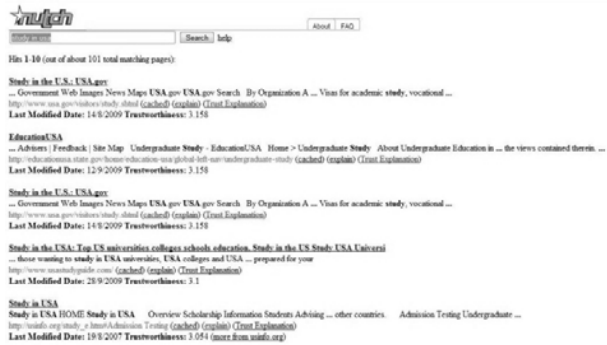

Fig.6. Modified NUTCH search results for the query "Study in US"

## 7. Conclusion and Future Work

The web search results from the search engines which implement the Trustworthy and High-Quality Information Retrieval System contain more accurate websites with trustworthy information. The search results provide the most truthful information. The search results are also ranked based on user-selected quality criteria. Performance of retrieving trustworthy data is also improved.

There are about 16 factors which affect the Content Trust of websites[5]. The future work will be in analyzing the remaining parameters and checking their feasibility in providing trustworthiness.

**Sumalatha Ramachandran** received the B.E. degree in Computer Science from Madras University and M.E. degree in Electronics Engineering from Madras Institute of Technology, Anna University. She has completed her Ph.D in the area of information retrieval system. She is currently working in the area of Semantic Information retrieval using visualization techniques, Database management systems and web services.

**Sujaya Paulraj, Sharon Joseph, and Vetriselvi Ramaraj** are members of the IEEE, IEEE Computer Society, ACM and students of Information technology, Madras Institute of Technology, Anna University.